\documentclass[a4paper,12pt]{article}
\pdfoutput=1 

\usepackage[utf8]{inputenc}
\usepackage{jheppub}
\usepackage{amsmath}
\usepackage{amsfonts}
\usepackage{amssymb}
\usepackage{mathtools}
\usepackage{amsthm}
\usepackage{dsfont}
\usepackage{hyperref}
\usepackage[capitalise,sort]{cleveref}
\usepackage{pgfplots}
\usepackage{tabularx}
\usepackage{environ}
\usepackage[inline]{enumitem}
\usepackage{comment}
\usepackage{tikz}
\usetikzlibrary{decorations.pathmorphing,calc,fit,matrix,cd,shapes,arrows,positioning, intersections,chains,positioning,arrows.meta,hobby,patterns,shadings,shadows}
\usepackage{blkarray}
\usepackage{shadow}
\usepackage{fancybox}
\usepackage{bm}
\usepackage{lscape}
\usepackage{verbatim}
\usepackage[normalem]{ulem}
\usepackage{subcaption}
\usepackage{mdframed}
\usepackage{orcidlink}

\pgfplotsset{
        compat=1.9,
        compat/bar nodes=1.8,
    }

\makeatletter
\def\@xfootnote[#1]{%
	\protected@xdef\@thefnmark{#1}%
	\@footnotemark\@footnotetext}
\makeatother

\definecolor{prhigh}{HTML}{ff0000}
\definecolor{sechigh}{HTML}{e0fbfc}
\definecolor{prcolor}{HTML}{1d3557}
\definecolor{seccolor}{HTML}{457b9d}
\definecolor{tercolor}{HTML}{98c1d9}
\definecolor{blueplot}{HTML}{58468e}

\newcommand\bea{\begin{eqnarray}}
\newcommand\eea{\end{eqnarray}}

\theoremstyle{plain}

\theoremstyle{definition}

\newtheorem*{conjecture*}{Conjecture}

\newtheorem{remark*}{Remark}



\makeatletter
\newcommand{\inlineitem}[1][]{%
\ifnum\enit@type=\tw@
    {\descriptionlabel{#1}}
  \hspace{0pt}%
\else
  \ifnum\enit@type=\z@
      \hspace{-15pt} \refstepcounter{\@listctr}\fi
    \quad\@itemlabel\hspace{0pt}%
\fi}
\makeatother
\DeclareMathAlphabet{\mathdutchcal}{U}{dutchcal}{m}{n}

\tikzset{
    partial ellipse/.style args={#1:#2:#3}{
        insert path={+ (#1:#3) arc (#1:#2:#3)}
    }
}

\tikzset{cross/.style={cross out, draw=black, fill=none, minimum size=2*(#1-\pgflinewidth), inner sep=0pt, outer sep=0pt}, cross/.default={2pt}}

\tikzset{
	pics/torus/.style n args={3}{
		code = {
			\providecolor{pgffillcolor}{rgb}{1,1,1}
			\begin{scope}[
				yscale=cos(#3),
				outer torus/.style = {draw,line width/.expanded={\the\dimexpr2\pgflinewidth+#2*2},line join=round},
				inner torus/.style = {draw=pgffillcolor,line width={#2*2}}
				]
				\draw[outer torus] circle(#1);\draw[inner torus] circle(#1);
				\draw[outer torus] (180:#1) arc (180:360:#1);\draw[inner torus,line cap=round] (180:#1) arc (180:360:#1);
			\end{scope}
		}
	}
}

\tikzset{
	pics/hole/.style n args={2}{
		code = {
			\draw[fill=white] (0,0) arc(120:60:#1 and #2)  arc(-60:-120:#1 and #2);
            \draw (0,0) arc(-120:-130:#1 and #2) (#1,0) arc(-60:-50:#1 and #2);
		}
	}
}

\makeatletter
\newcommand*{\itemequation}[3][]{%
  \item
  \begingroup
    \refstepcounter{equation}%
    \ifx\\#1\\%
    \else  
      \label{#1}%
    \fi
    \sbox0{#2}%
    \sbox2{$\displaystyle#3\m@th$}%
    \sbox4{\@eqnnum}%
    \dimen@=.5\dimexpr\linewidth-\wd2\relax
    \ifcase
        \ifdim\wd0>\dimen@
          \z@
        \else
          \ifdim\wd4>\dimen@
            \z@
          \else 
            \@ne
          \fi 
        \fi
      \@latex@warning{Equation is too large}%
    \fi
    \noindent   
    \rlap{\copy0}%
    \rlap{\hbox to \linewidth{\hfill\copy2\hfill}}%
    \hbox to \linewidth{\hfill\copy4}%
    \hspace{0pt}
  \endgroup
  \ignorespaces 
}
\makeatother  

\hypersetup{
	pdftitle={Formulating the Weak Gravity Conjecture in AdS Space},    
	pdfauthor={\textcopyright\ Puxin Lin, Alessandro Mininno, Gary Shiu},     
	pdfsubject={hep-th},   
	pdfcreator={pdfLaTex},   
	pdfproducer={LaTex}, 
	pdfkeywords={},
	colorlinks=true
  ,urlcolor=blue
  ,anchorcolor=blue
  ,citecolor=blue
  ,filecolor=blue
  ,linkcolor=blue
  ,menucolor=blue
  ,linktocpage=true
}

\allowdisplaybreaks[1] 

\crefname{figure}{Figure}{Figures}
\crefname{table}{Table}{Tables}
\crefname{definition}{Definition}{Definitions}
\crefname{proposition}{Proposition}{Propositions}
\crefname{claim}{Claim}{Claims}
\crefname{conjecture}{Conjecture}{Conjectures}

\def\beq{\begin{equation}}
\def\eeq{\end{equation}}
\def\bc{\begin{cases}}
\def\ec{\end{cases}}
\def\bal{\begin{aligned}}
\def\eal{\end{aligned}}

\arxivnumber{}
\title{Quantum Corrections and Extremality: A Generalized Universal Relation}
\author{Ankit Anand\,\orcidlink{0000-0002-8832-3212}}

\affiliation{Department of Physics, Indian Institute of Technology, Kanpur 208016, India}

\emailAdd{anand@iitk.ac.in}

\abstract{Logarithmic corrections to the entropy of extremal black holes have proven effective in precisely matching the microscopic degeneracies obtained from string-theoretic as well as a non-perturbative quantum correction manifests as an exponential term in the black hole entropy. In this work, we extend the universal relation proposed by Goon and Penco by deriving a generalized form where entropy is not just the Bekenstein-Hawking entropy. Our analysis treats entropy as a general function of the horizon radius, and with the help of that, we formulate the generalized universal relation. We show that, in the case of Bekenstein-Hawking entropy, the generalized relation coincides with the original universal relation by Goon and Penco. Furthermore, we explore the implications of logarithmic and exponential corrections to entropy and test the validity of the generalized universal relation under these modifications.
}
\begin{document}

\maketitle

\section{Introduction}\label{Sec: Introduction}

Quantum black hole entropy constitutes the full quantum refinement of the renowned Bekenstein–Hawking entropy \cite{Bekenstein:1973ur, Bekenstein:1974ax}, encompassing both perturbative expansions and non-perturbative contributions arising from quantum gravitational effects. A fundamental test for any candidate theory of quantum gravity lies in its ability to reconcile the macroscopic and microscopic descriptions of black hole entropy. This comparison serves as a stringent consistency check and reveals a profound connection between the infrared (IR) and ultraviolet (UV) regimes of gravitational dynamics. The leading-order contribution to black hole entropy, encapsulated by the Bekenstein–Hawking area law, arises from the theory's low-energy (IR) sector. However, it must precisely match the microscopic entropy obtained by counting the degeneracy of quantum microstates, a quantity inherently sensitive to the UV completion of the theory. This dual requirement exemplifies the holographic nature of gravity and highlights the necessity for a unified framework wherein both descriptions coexist consistently. Significant progress has been made in this direction through approaches such as string theory, loop quantum gravity, and the AdS/CFT correspondence. In particular, black hole solutions in anti-de Sitter (AdS) spacetimes offer a robust setting for exploring quantum aspects of gravity and for testing holographic dualities. Beyond the leading Bekenstein–Hawking term, subleading corrections—especially logarithmic terms—encode essential information about quantum fluctuations and the statistical mechanics of the underlying microstates.

\quad The logarithmic term is especially significant, as it arises generically from one-loop quantum fluctuations of fields propagating in the black hole background and has been computed across various approaches, including Euclidean path integrals, conformal field theory techniques, and holographic renormalization group flows \cite{Dijkgraaf:1996it, Dabholkar:2008zy, Mandal:2010cj, Ashtekar:1997yu, Kaul:2000kf, Meissner:2004ju, Domagala:2004jt, Ghosh:2004wq, Ghosh:2012jf, Modak:2025gvp}. For black holes with horizon areas approaching the Planck scale \( A \sim \mathcal{O}(\ell_P^2) \), these corrections may become dominant, and the semiclassical expansion itself may break down. In such regimes, a full nonperturbative quantum gravitational description becomes essential, and the precise form of entropy is expected to deviate substantially from the semiclassical expansion. In contrast, exponential corrections to black hole entropy have not been investigated in the literature with the same depth. Although certain computations in string theory suggest the presence of such corrections \cite{Dabholkar:2014ema}. Recently, \cite{Chatterjee:2020iuf} derived the black hole entropy and identified an exponential correction term using only the intrinsic geometry of the horizon, without recourse to string theory or loop quantum gravity. Their analysis proceeds by identifying local microstates associated with the black hole horizon and leads to a discrete area spectrum. These results indicate that exponential corrections may naturally emerge in a broad class of quantum gravity theories and could represent a universal feature of horizon microstate quantization. Quantum gravitational effects introduce corrections to the classical Bekenstein–Hawking entropy, particularly for black hole horizons with areas much larger than the Planck scale \( \ell_P^2 \). As suggested in \cite{Chatterjee:2020iuf}, one should accept that the entropy of a black hole admits a quantum-corrected expansion of the form
\begin{eqnarray}\label{Quantum Entropy Expansion}
S &=& \frac{A}{4\ell_P^2} + \alpha \ln\left(\frac{A}{4\ell_P^2}\right) + \frac{\beta}{4\ell_P^2 A} + \cdots + \exp\left(-\delta \frac{A}{4\ell_P^2}\right) + \cdots \ , \\
&\sim& \frac{r_h^2}{\ell_P^2} + \alpha \ln\left(\frac{r_h^2}{\ell_P^2}\right) + \frac{\beta}{\ell_P^2 r_h^2} + \cdots + \exp\left(-\delta \frac{r_h^2}{\ell_P^2}\right) + \cdots \ , \nonumber
\end{eqnarray}
where $r_h$ is the horizon radius and \( \alpha \), \( \beta \), \( \delta \), etc., are universal coefficients determined by the underlying theory of quantum gravity. These corrections include both logarithmic and inverse-area terms and exponentially suppressed contributions, reflecting perturbative and non-perturbative quantum effects beyond the semiclassical regime.

\quad Perturbative corrections to black hole entropy can also emerge from thermal fluctuations \cite{Das:2001ic, Upadhyay:2017fiw, Pourhassan:2019bub, Sadeghi:2016dvc}. Within the Jacobson formalism, where spacetime geometry is viewed as an emergent thermodynamic construct \cite{Jacobson:1995ab}, these thermal fluctuations are directly related to quantum fluctuations, and using that correspondence, one can derive the quantum corrections to the geometry of various black holes via thermodynamic fluctuations \cite{Faizal:2017drd, Hammad:2015ipa, Pourhassan:2019coq, Pourhassan:2017qhq, Pourhassan:2017qxi}. For large black holes, quantum and thermal fluctuations are typically negligible. However, as the black hole shrinks, both effects become significant due to rising temperature and decreasing size. At intermediate scales, these corrections are well-described by perturbative expansions, but near the Planck scale, non-perturbative effects dominate and must be incorporated. Such corrections often appear as exponential functions of the classical entropy, and have been obtained using number-theoretic structures like Kloosterman sums \cite{Dabholkar:2014ema} as well as through AdS/CFT correspondence involving massless supergravity fields near the horizon \cite{Duff:1994an, Peet:2000hn}. These non-perturbative contributions, rooted in string-theoretic frameworks, are particularly relevant for extended objects such as black branes \cite{Zhou:2015yxa, Lu:2013nt, Lu:2012rm, Xiao:2015bha}.

\quad Goon and Penco \cite{Goon:2019faz} examined the thermodynamic behavior of near-extremal black holes under perturbative corrections. They demonstrated that the free energy yield perturbations modified thermodynamic relations among mass, entropy, and temperature and satisfy the universal relation
\begin{equation}\label{Universal Relation}
   -\,\frac{\left(\partial M_{ext}/\partial\varepsilon \right)}{ T \left(\partial S/\partial\varepsilon \right)|_{M_{ext}}} = 1 \ ,
\end{equation}
where \( M \), and \( \varepsilon \) denote mass and the perturbative parameter, respectively\footnote{In the original paper of Goon and Penco \cite{Goon:2019faz}, the denominator is $ \lim_{M \rightarrow M_{ext}} -T\left(\frac{\partial S(M,\vec{\mathcal{Q}},\varepsilon)}{\partial \varepsilon}\right)$, in this paper, we will use this expression as $-T \left(\partial S/\partial\varepsilon \right)|_{M_{ext}}$}. The $M_{ext}$ and $S$ also depend on other hairs of the black hole, and those are suppressed here. In leading order, these corrections can be interpreted in terms of higher-derivative contributions to the effective action \cite{Wang:2022sbp}. Subsequent developments have extended this framework to various asymptotically AdS geometries, illustrating the relevance of the entropy–extremality relation \cite{Cremonini:2019wdk, Cano:2019oma, Sadeghi:2020xtc, Wei:2020bgk, Chen:2020rov, Chen:2020bvv, Sadeghi:2020ciy, Ma:2020xwi, McPeak:2021tvu, McInnes:2021frb, Etheredge:2022rfl, Sadeghi:2022xcr, Ko:2023nim, Anand:2024wxa, Anand:2024pze, Lee:2024thc, Ko:2024ymy, Anand:2024wjr, Anand:2024hpj, Khosravani:2024xvz, Sadeghi:2020ntn, ANAND2025101916}. A key result is the emergence of proportionality between shifts in entropy and mass, where a negative proportionality constant ensures the WGC remains satisfied \cite{Cheung:2019cwi, Kats:2006xp, Reall:2019sah, Cheung:2018cwt}. Within the framework of quantum gravity, the Weak Gravity Conjecture (WGC), initially proposed by Vafa \cite{Vafa:2005ui, Arkani-Hamed:2006emk}, provides a criterion for the dominance of gauge interactions over gravitational attraction. The conjecture stipulates that for a consistent theory incorporating a U(1) gauge symmetry, a state must exist, typically associated with an extremal black hole, for which the charge-to-mass ratio satisfies the inequality \(Q\geq M \). This bound ensures that gravity is not the strongest force in the spectrum and plays a crucial role in precluding exact global symmetries in quantum gravity theories. A compelling argument supporting the WGC arises from the behavior of black hole evaporation. As Hawking radiation is largely insensitive to global charges, black holes can radiate without preserving global quantum numbers, thereby aligning with the WGC’s exclusion of global symmetries \cite{Banks:2010zn, Harlow:2022ich}. Multiple approaches have been employed to substantiate the WGC, including analyses of higher-derivative corrections aimed at circumventing the formation of naked singularities. The prevention of such singularities often indirectly verifies the WGC bound \cite{Cheung:2018cwt}.

\quad The motivation for this work is that numerous studies have investigated the validity of the universal relation proposed by Goon and Penco in various gravitational and black hole settings. While several works support the relation under specific conditions, reported deviations indicate that the universality may not be preserved \cite{Sadeghi:2022xcr} in certain black hole solutions, mainly when the entropy is not the Bekenstein-Hawking entropy. Motivated by these findings, we consider a generalized entropy functional in this paper and examine its implications for the universal relation within the context of Logarithmic and Exponential corrected entropy. This approach allows us to explore the conditions under which the relation remains valid or breaks down. We derive a generalized universal relation based on \cite{Goon:2019faz} connecting the leading-order quantum or higher-derivative corrections to both the extremality bound and the generalized entropy\footnote{The generalized entropy stands the entropy is not just the Bekenstein–Hawking entropy; it is a function of the horizon radius $r_h$ i.e., $S = \tilde{\mathrm{f}}\,(r_h)$. For the Bekenstein–Hawking entropy $\tilde{\mathrm{f}}\,(r_h) = \pi r_h^2$} as 
\begin{equation}\label{Generalized Universal Relation}
  \boxed{  - \, \frac{(\partial M_{ext}/\partial \varepsilon)}{T\,(\partial S/\partial \varepsilon)|_{M_{ext}}}=  \frac{\partial S_{BH}^h}{\partial S}}
\end{equation} 
where $S_{BH}^h =\pi \,\underline{\mathrm{r_h}}^2$ is the Bekenstein–Hawking entropy but for the horizon radius ($\underline{\mathrm{r_h}}$) derived from Eq.~\eqref{Quantum Entropy Expansion}\footnote{As described in Eq.\eqref{Quantum Entropy Expansion}, the entropy is
\begin{eqnarray}
    S=\tilde{\mathrm{f}}(r_h) \xrightarrow{\text{by inverting them }}r_h = \tilde{\mathrm{f}}^{-1}(S)=r_h(S) \ .
\end{eqnarray}
For convenience, throughout this work, we denote $r_h(S)=\underline{\mathrm{r_h}}.$ Also $\mathrm{r_h}'=\frac{\partial \underline{\mathrm{r_h}}}{\partial S}$.}. We consider a generic gravitational theory admitting stationary black holes parametrized by the ADM mass \( M \) and the other conserved charges such as gauge charges, angular momenta, or other global quantum numbers. The classical thermodynamic description of these solutions is encoded in the entropy function \( S_0 \) and corresponding temperature \( T_0\). For fixed charge configurations, physical black hole states are constrained by an extremality bound of the form \( M > M_{\rm ext}^{0} \), where \( M_{\rm ext}^{0} \) denotes the classical extremal mass threshold below which no regular black hole solutions exist. We consider a perturbative deformation of the gravitational theory governed by an expansion parameter \( \varepsilon \). This deformation induces corrections to the thermodynamic quantities associated with black hole solutions, rendering both the extremality bound and entropy functions explicitly \( \varepsilon \)-dependent. Accordingly, the extremality condition takes the form \( M > M_{\rm ext}(\varepsilon) \), while the entropy generalizes to \( S \), with the requirement that the undeformed quantities are recovered in the limit \( \varepsilon \to 0 \) 
\begin{equation}
M_{\rm ext}= M_{\rm ext}^{0} \;\;\;\;\;\;\;\;\; \text{and} \;\;\;\;\;\;\;\;\; S = S_0 \ .
\end{equation}
Under a minimal set of assumptions—chiefly, the existence of a consistent thermodynamic description and a smooth deformation parameter \( \varepsilon \)—we show that the universal relation \eqref{Universal Relation} will not hold, and we derive the generalized universal relation and show that for the Bekenstein–Hawking entropy the generalized universal relation \eqref{Generalized Universal Relation} is same as the universal relation \eqref{Universal Relation}.


\quad The paper is organized as follows: In Section \ref{Sec:Review of Universal Relation}, we briefly review the universal relation proposed by Goon and Penco and then discuss how quantum corrections can modify the entropy, introducing exponential or logarithmic terms, without affecting the classical background geometry. In Section \ref{Sec:Generalized Universal Relation}, we derive the central result of this work: a Generalized Universal Relation \eqref{Generalized Universal Relation}, which we show is exactly satisfied only when the entropy assumes the Bekenstein–Hawking form. This establishes the area law as uniquely compatible with the underlying universality structure. In Section \ref{Sec:Logarithmic correction}, we analyze generic logarithmic corrections as typically arising from one-loop quantum effects and demonstrate that the universal relation persists in this context. In Section \ref{Sec:Exponential COrrection}, we consider general exponential corrections motivated by effective field theory and modified gravity models and verify that the generalized relation remains valid. Section \ref{Sec:Conclusion and Discussion} summarizes results and discusses their relevance to the microscopic structure of black hole entropy and the imprint of quantum gravitational effects on thermodynamic consistency conditions.

\section{Review of Universal Relation}\label{Sec:Review of Universal Relation}

This section briefly discusses the universal relation as discussed in \cite{Goon:2019faz}. We start with the Reissner–Nordström–Anti-de Sitter black hole, which arises as a static, spherically symmetric solution to the Einstein-Maxwell action with a negative cosmological constant. The action governing the dynamics of the spacetime and gauge fields is given by  
\begin{equation} \label{Action}
\mathcal{I} = \frac{1}{16\pi} \int d^4x \sqrt{-g} \left( R - F_{\mu\nu}F^{\mu\nu} + \frac{6}{l^2} \right) \ ,
\end{equation}
where \( R \) is the Ricci scalar, \( F_{\mu\nu}= \partial_\mu A_\nu - \partial_\nu A_\mu \) is the electromagnetic field strength tensor associated with a \( U(1) \) gauge field \( A_\mu \), and \( l \) denotes the AdS curvature radius, related to the cosmological constant via \( \Lambda = -3/l^2 \). The resulting field equations admit a solution with the line element
\begin{equation}
ds^2 = -f(r)\, dt^2 + \frac{dr^2}{f(r)} + r^2 \left( d\theta^2 + \sin^2\theta\, d\phi^2 \right) \ .
\end{equation}
The metric function \( f(r) \), determined by solving the Einstein-Maxwell equations, takes the explicit form
\begin{equation}\label{Charged Metric Fuction}
f(r) = 1 - \frac{2M}{r} + \frac{Q^2}{r^2} + \frac{r^2}{l^2} \ ,
\end{equation}
where \( M \) denotes the ADM mass and $Q$ is the electric charge of the black hole. The black hole's mass by solving $f(r=r_h=0)$ and temperature by using $\frac{1}{4 \pi}\frac{d\,f(r)}{d\,r}\Big|_{r=r_h}$. The entropy of the black hole is 
\begin{equation*}
    S=\pi \, r_h^2 \ ,
\end{equation*}
where $r_h$ is the outer event horizon's radius. Now, we consider a perturbative correction to the original gravitational action given by  
\begin{equation}\label{Perturbation Action}
    \Delta \mathcal{I} = \frac{3}{8\pi} \int d^{4}x\, \sqrt{-g} \, \frac{ \varepsilon}{l^2} \ ,
\end{equation}
where \( \varepsilon \ll 1 \) is a small expansion parameter and reduces to the original uncorrected action in the limit \( \varepsilon \to 0 \). The total action can be written as
\begin{equation}
    \mathcal{I}_{\text{Tot.}}= \mathcal{I} + \Delta\mathcal{I} \ .
\end{equation}
Due to this correction, the resulting black hole solution acquires corrections that manifest as shifts in the metric functions as 
\begin{equation}
    f_{\text{Tot.}}(r,\varepsilon) = f(r) +\varepsilon \frac{r^2}{l^2} +\mathcal{O}(\varepsilon^2) \ .
\end{equation}
These corrections, in turn, also modify the black hole’s thermodynamic properties. In particular, the mass and the Hawking temperature receive a perturbative correction and are accordingly modified from their original form. The modified thermodynamic quantities 
\begin{eqnarray}\label{RN Perturbed MAss and Temperature}
    M(\varepsilon) =\frac{\pi  l^2 \left(\pi  Q^2+S\right)+S^2 (\varepsilon +1)}{2 \pi ^{3/2} l^2 \sqrt{S}} \;\;\;\;\;\;\;;\;\;\;\;\;\;\; T(\varepsilon) =\frac{\pi  l^2 \left(S-\pi  Q^2\right)+3 S^2 (\varepsilon +1)}{4 \pi ^{3/2} l^2 S^{3/2}} \ .
\end{eqnarray}
Using Eq.~\eqref{RN Perturbed MAss and Temperature}, one can easily find 
\begin{equation}\label{RN Epsilon}
    \varepsilon =  \frac{\pi  l^2 \left(2M \sqrt{\pi \,S}-\pi  Q^2-S\right)}{S^2}-1 \ .
\end{equation}
To test the universal relation as in Eq.~\eqref{Universal Relation}, we can easily compute 
\begin{eqnarray}
    \frac{\partial M_{ext}}{\partial \varepsilon} &=& \frac{S^{3/2}}{2 \pi ^{3/2} l^2} \\
    T(\varepsilon)\left(\frac{\partial S}{\partial \varepsilon}\right)\Bigg|_{M_{ext}} &=& \frac{S^{3/2} \left(\pi  l^2 \left(S-\pi  Q^2\right)+3 S^2 (\varepsilon +1)\right)}{4 \pi ^{5/2} l^4 \left(-3 \sqrt{\pi } M \sqrt{S}+2 \pi  Q^2+S\right)}
\end{eqnarray}
Finally, using Eq.~\eqref{RN Epsilon}, we have verified the universal relation \eqref{Universal Relation}. One can consider the simplest case and check whether the universal relation is satisfied. In Appendix \ref{Appn:rh^a Entropy}, we have studied the universal relation for the simplest entropy relation\footnote{This relation we have assumed just to show that if $\mathrm{a}\neq 2$ the universal relation \eqref{Universal Relation} is not satisfied.} 
\begin{eqnarray}
    S = \pi r_h^\mathrm{a}
\end{eqnarray}
It is obvious to check that the generalized universal relation \eqref{Generalized Universal Relation} can be satisfied using this form of entropy.

\quad The Bekenstein-Hawking entropy of a black hole can receive both logarithmic and exponential quantum corrections while maintaining the classical spacetime geometry. Such corrections naturally arise in various quantum gravity frameworks, including loop quantum gravity, string theory, and effective field theory, and are attributed to quantum fluctuations around the classical background rather than modifications to the classical equations of motion. However, quantum gravitational effects induce subleading corrections to this entropy, which can be systematically analyzed within various theoretical frameworks.

\quad Logarithmic Corrections: Among these quantum corrections, logarithmic terms are particularly robust and universal across different approaches to quantum gravity. In the context of loop quantum gravity, the quantization of geometric operators leads to a discrete spectrum for the horizon area. The microstate counting in this framework yields an entropy that, beyond the leading order, includes a correction proportional to the logarithm of the horizon area:
\[
\sim \alpha \ln \left(\frac{A}{G}\right) \ ,
\]
where the coefficient \(\alpha\) depends on the specific quantum gravity model, such as the number of massless fields or the details of the spin network states in loop quantum gravity. Similarly, in string theory, microscopic calculations for extremal and near-extremal black holes—often performed via conformal field theory techniques and the Cardy formula—reproduce the leading entropy and include subleading logarithmic corrections arising from finite-size effects, higher-genus contributions, and quantum fluctuations of the microstates. These corrections are also captured within the Euclidean path integral formalism, where quantum fluctuations of matter and gravitational fields around the classical background contribute to the partition function, resulting in a correction term proportional to \(\ln(A/G)\). Importantly, these logarithmic corrections do not alter the classical metric itself; instead, they reflect the quantum and microscopic degrees of freedom that modify the entropy without changing the classical solution.

\quad Exponential Corrections : Beyond the logarithmic terms, the entropy may also receive exponentially suppressed contributions from non-perturbative quantum gravity effects. Such effects include instanton configurations or other non-perturbative phenomena in the Euclidean gravitational path integral, leading to corrections of the form
\[
\sim e^{-A/G} \ .
\]
These exponential terms are generally negligible for large black holes but are crucial for understanding the full quantum structure of black hole microstates. In the holographic context, particularly within the AdS/CFT correspondence, these corrections can be interpreted as \(1/N\)-suppressed effects in the dual conformal field theory, corresponding to quantum corrections beyond the planar limit. Although less universal than the logarithmic corrections, the presence of exponential terms is consistent with the expectation that the full quantum gravity partition function encompasses non-perturbative contributions. Crucially, these non-perturbative corrections do not necessitate modifications to the classical metric; they are quantum effects encoded in the path integral and microstate counting.

\section{Generalized Universal Relation}\label{Sec:Generalized Universal Relation}

In this Section, we derive the generalized universality relation. The derivation starts with assuming that the entropy is not the Bekenstein–Hawking entropy. We start with the general expression of entropy as a function of entropy similar to Eq.~\eqref{Quantum Entropy Expansion}, we start by assuming that the inverse function exists. This assumption allows us to express the horizon radius as a function of entropy. Thus, we can write
\begin{equation}\label{General Horizon Radius}
    r_h = \tilde{\mathrm{f}}^{-1}(S) = r_h(S) \ .
\end{equation}
For notational convenience, throughout this work we denote the function $r_h(S)$ simply by $\underline{\mathrm{r_h}}$, with the understanding that $r_h(S) = \underline{\mathrm{r_h}}$ unless stated otherwise as well as $\underline{\mathrm{r_h}}'$ represents the differentiation of $r_h(S)$ w.r.t. $S$. Using this expression and substituting it into the black hole solution characterized by the metric function in Eq.~\eqref{Charged Metric Fuction}, we can now compute the Mass and the Hawking temperature in terms of the entropy function \( r_h(S) \). We perturb the system by introducing a small parameter \( \varepsilon \) to study quantum corrections or any small deviation from the classical theory as discussed in Sec.~\ref{Sec:Review of Universal Relation}. The corrected thermodynamic quantities, up to first order in \( \varepsilon \), take the form:
\begin{eqnarray}\label{General Perturbed Quantities}
    M = \frac{(\varepsilon +1) \underline{\mathrm{r_h}}^4+l^2 \underline{\mathrm{r_h}}^2+l^2 Q^2}{2 l^2\underline{\mathrm{r_h}}} \;\;\;\;\;\;\;\;;\;\;\;\;\;\;\;\;
    T = \frac{l^2 \underline{\mathrm{r_h}}^2+3 (\varepsilon +1) \underline{\mathrm{r_h}}^4-l^2 Q^2}{4 \pi  l^2 \underline{\mathrm{r_h}}^3} \ .
\end{eqnarray}
By comparing the perturbed and unperturbed mass expressions, we can isolate the perturbation parameter \( \varepsilon \) as
\begin{equation}\label{Epsilon}
    \varepsilon = \frac{2 l^2 M}{\underline{\mathrm{r_h}}^3}-\frac{l^2 Q^2}{\underline{\mathrm{r_h}}^4}-\frac{l^2}{\underline{\mathrm{r_h}}^2}-1 \ .
\end{equation}
Using them, we can easily compute the numerator and denominator of Eq.~\eqref{Generalized Universal Relation} as 
\begin{equation}
    -T \left(\frac{\partial S}{\partial \varepsilon}\right)\Bigg|_{M_{ext}} = \frac{\underline{\mathrm{r_h}}^2}{4 \, \pi \, l^2 \,\underline{\mathrm{r_h}}'}\;\;\;\;\;\;\;\;\text{and}\;\;\;\;\;\;\;\;\; \frac{\partial M_{ext}}{\partial \varepsilon}=\frac{\underline{\mathrm{r_h}}^3}{2 l^2} \ .
\end{equation}
From this, it is easy to check that the generalized universality relation \eqref{Generalized Universal Relation} is satisfied. The universal thermodynamic relation, as in \eqref {Universal Relation}, is satisfied if and only if the following condition holds
\begin{equation}
   \frac{d\underline{\mathrm{r_h}}}{dS}  = \frac{1}{2 \pi \underline{\mathrm{r_h}}} \ .
\end{equation}
Solving this differential equation yields a unique functional form
\begin{eqnarray}
    r_h(S) = \sqrt{\frac{S}{\pi} + S_0} \ ,
\end{eqnarray}
where \( S_0 \) is an integration constant. Setting \( S_0 = 0 \) retrieves the well-known Bekenstein–Hawking area law \( S = \pi r_h^2 \), implying that only this form of entropy strictly satisfies the universal relation. Any deviation from this entropy form leads to violations of the relation, highlighting the special role of the classical Bekenstein–Hawking entropy in the thermodynamic consistency of black hole physics.  Before closing this section, we can also comment on other universality relations as discussed in \cite{Chen:2020bvv}. We start by computing 
\begin{eqnarray}\label{D P and DQ by epsilon}
    \frac{\partial Q}{\partial \varepsilon} = -\frac{\underline{\mathrm{r_h}}^4}{2 l^2 Q} \;\;\;\;\;\;\;\;;\;\;\;\;\;\;\;\; \frac{\partial P}{\partial \varepsilon} = -\frac{3}{8 \pi  l^2} \ ,
\end{eqnarray}
where the cosmological constant was treated as a fixed parameter. However, in the context of extended black hole thermodynamics, it is now reinterpreted as a dynamical variable associated with the thermodynamic pressure. Specifically, the pressure in terms of AdS curvature radius is defined as $8\pi P= -\Lambda$. The conjugate thermodynamic quantity corresponding to this pressure is the thermodynamic volume \( V \). Using Eq.~\eqref{D P and DQ by epsilon}, we can verify that 
\begin{equation}
    \frac{\partial M_{ext}}{\partial \varepsilon} = - \Phi \left( \frac{\partial Q}{\partial \varepsilon}\right)_{M{ext}} = -V  \left(\frac{\partial P}{\partial \varepsilon}\right)_{M_{ext}} \ ,
\end{equation}
where $\Phi$ is potential conjugate to charge $Q$. Now, we will test the result in the known theory where entropy gets modified by a logarithmic term in the Einstein–Gauss–Bonnet theory.

\subsection{An Example of Charged black hole in EGB Gravity}

In this section, we first revisit the charged Anti-de Sitter black hole solution within the context of the specified gravitational framework. We then perform a perturbation to the metric as discussed in Sec.~\ref{Sec:Review of Universal Relation} and proceed with a similar computation of Sec.~\ref {Sec:Generalized Universal Relation}.

\quad We consider the action of a $\mathbf{D}$-dimensional Einstein–Gauss-Bonnet (EGB) gravity theory minimally coupled to a Maxwell field, where the Gauss-Bonnet coupling constant $\alpha$ is rescaled as $\alpha/(D-4)$ to enable a smooth extension to four dimensions. The total action is given by  
\begin{equation}
\mathcal{I} = \frac{1}{16\pi} \int d^D x\, \sqrt{-g} \left( R - 2\Lambda + \frac{\alpha}{D-4} \mathcal{L}_{GB} - F_{\mu\nu} F^{\mu\nu} \right),  
\label{acI}
\end{equation}
where $R$ is the Ricci scalar, $\Lambda$ denotes the cosmological constant, and $F_{\mu\nu} = \partial_\mu A_\nu - \partial_\nu A_\mu$ is the electromagnetic field strength tensor associated with the gauge potential $A_\mu$. The Gauss-Bonnet Lagrangian $\mathcal{L}_{GB}$ captures the second-order curvature corrections and is defined as
\begin{equation}
\mathcal{L}_{GB} = R_{\mu\nu\lambda\rho} R^{\mu\nu\lambda\rho} - 4 R_{\mu\nu} R^{\mu\nu} + R^2,
\end{equation}
which arises naturally in the low-energy limit of heterotic string theory and provides ghost-free higher-curvature modifications in $D \geq 5$. Varying the action \eqref{acI} with respect to the metric $g_{\mu\nu}$ and the gauge field $A_\mu$ yields the coupled field equations governing the gravitational and electromagnetic sectors
\begin{align}
G_{\mu\nu} - \Lambda g_{\mu\nu} + \frac{\alpha}{D-4} H_{\mu\nu} = T_{\mu\nu} \;\;\;\;\;\;\;\;\;;\;\;\;\;\;\;
\nabla_\mu F^{\mu\nu} = 0 \ , 
\end{align}
where $G_{\mu\nu}$ denotes the Einstein tensor, $H_{\mu\nu}$ is the Lanczos tensor associated with the variation of the Gauss-Bonnet term, and $T_{\mu\nu}$ is the energy-momentum tensor of the Maxwell field. These tensors are explicitly given by
\begin{align}
H_{\mu\nu} &= 2 \left( R R_{\mu\nu} - 2 R_{\mu\sigma} R^\sigma_{~\nu} - 2 R_{\mu\sigma\nu\rho} R^{\sigma\rho} - R_{\mu\sigma\rho\lambda} R_{\nu}^{~\sigma\rho\lambda} \right) - \frac{1}{2} g_{\mu\nu} \mathcal{L}_{GB}, \\
T_{\mu\nu} &= 2 F_{\mu}^{~\lambda} F_{\nu\lambda} - \frac{1}{2} g_{\mu\nu} F^{\alpha\beta} F_{\alpha\beta}.
\end{align}

\quad We consider a class of four-dimensional, static, and spherically symmetric black hole spacetimes described by the line element  
\begin{equation}
ds^{2} = -f(r)\, dt^{2} + \frac{dr^{2}}{f(r)} + r^{2} \left( d\theta^{2} + \sin^{2}\theta\, d\phi^{2} \right),  
\label{Metric}
\end{equation}  
where the function \( f(r) \) encapsulates the gravitational potential and encodes the geometric and causal structure of the spacetime. To incorporate electromagnetic interactions, we adopt a purely electric Maxwell field through the gauge potential ansatz \( A_\mu = h(r) \delta^0_\mu \), consistent with the spacetime symmetries. Solving the Maxwell equations derived from the action functional yields the electrostatic potential as $h(r) = -\frac{Q}{r}$ where \( Q \) denotes the total electric charge of the black hole configuration \cite{Hegde:2020cdm, Fernandes:2020rpa}. We then seek solutions to the gravitational field equations arising from the Einstein–Gauss–Bonnet–Maxwell action in arbitrary spacetime dimensions \( D \), rescaled via the prescription \( \alpha \rightarrow \alpha/(D-4) \) to ensure the existence of a well-defined four-dimensional limit. This regularization circumvents the topological nature of the Gauss-Bonnet term in \( D=4 \) and yields non-trivial contributions to the gravitational dynamics. Solving the field equations in this setup and subsequently taking the limit \( D \rightarrow 4 \) leads to the static, spherically symmetric charged AdS black hole solution in four-dimensional EGB gravity, expressed as  
\begin{equation}
f(r) = 1 + \frac{r^2}{2\alpha} \left( 1 \pm \sqrt{1 + 4\alpha \left( \frac{2M}{r^3} - \frac{Q^2}{r^4} + \frac{\Lambda}{3} \right)} \right),  
\label{Eqmetric1}
\end{equation}  
where \( M \) and \( Q \) are interpreted as the ADM mass and electric charge of the black hole, \( \alpha \) is the Gauss-Bonnet coupling parameter, and \( \Lambda \) denotes the cosmological constant associated with AdS asymptotics. The solution possesses two distinct branches corresponding to the \( \pm \) sign. The positive branch is generally associated with pathological behavior such as ghost-like excitations or causality violation in perturbative analyses \cite{Boulware:1985wk}, and thus is deemed physically inadmissible. In the subsequent analysis, we confine ourselves to the negative branch, which yields a ghost-free, well-behaved gravitational solution in the low-energy limit. The mass of the black hole can be obtained by evaluating the metric function at the event horizon \( r = r_h \), where the condition \( f(r_h) = 0 \) must be satisfied. Solving this constraint leads to the following expression for the black hole mass:
\begin{equation}
M_{\text{EGB}} = \frac{r_h}{2} \left( 1 - \frac{\Lambda r_h^{2}}{3} + \frac{Q^{2} + \alpha}{r_h^{2}} \right),  
\label{Mass}
\end{equation}
where \( Q \) and \( \alpha \) denote the black hole charge and the Gauss-Bonnet coupling constant, respectively, while \( \Lambda \) represents the cosmological constant as well as the temperature of the black hole is 
\begin{equation}\label{T EGB}
    T_{\text{EGB}} = \frac{1}{4 \pi  r_h}-\frac{Q^2+\alpha}{4 \pi  r_h^3}+\frac{3 r_h}{4 \pi  l^2}
\end{equation}
Invoking the first law of black hole thermodynamics in the differential form \( dM = T\, dS \) and substituting the explicit forms of \( M \) and the Hawking temperature \( T \), the black hole entropy can be computed via the integral
\begin{equation}\label{SEGB}
S = \int_{0}^{r_h} \frac{dM}{T} = \pi r_h^2 + 4\pi \alpha \ln \left( \frac{r_h}{\ell_0} \right),  
\end{equation}
where \( \ell_0 \) is an arbitrary constant with dimensions of length introduced to render the argument of the logarithmic term dimensionless.

\quad The entropy expression above extends the Bekenstein–Hawking area law by incorporating a logarithmic correction term originating from the inclusion of higher-curvature Gauss–Bonnet corrections in the gravitational action. Such corrections are ubiquitous in various approaches to quantum gravity, including string theory and effective field theory formulations of gravitational dynamics, and they encode subleading contributions from quantum microstates of the black hole. The logarithmic term is particularly significant in regimes where the curvature becomes strong and quantum corrections become non-negligible. Now, we will delve into the calculation details by representing the horizon radius in terms of entropy as in Eq.~\eqref{SEGB}, we have
\begin{equation}\label{EGB Horizon}
    r_h^2 = 2 \alpha \, W\left(\frac{\ell_0^2\, e^{\frac{S}{2 \, \pi \, \alpha }}}{2 \alpha }\right) = 2 \alpha \, \mathcal{W} \ .
\end{equation}
Here $W\left(\frac{\ell_0^2\, e^{\frac{S}{2 \, \pi \, \alpha }}}{2 \alpha }\right)$ is the Lambert W function detail discussion in Appendix \ref{Appn:Lambert W function} also we have defined $\mathcal{W} = W\left(\frac{\ell_0^2\, e^{\frac{S}{2 \, \pi \, \alpha }}}{2 \alpha }\right)$. By adding the perturbation in the metric similar to \eqref{Perturbation Action}, the metric gets modified, and then the thermodynamic quantities as well. Using Eq.\eqref{Mass},\eqref{T EGB} and \eqref{EGB Horizon}, the corrected mass and temperature of the black hole are 
\begin{eqnarray}\label{Perturbed EGB}
    M = \frac{2 \alpha  \mathcal{W}\left(l^2+2 \alpha  (\varepsilon +1) \mathcal{W}\right)+l^2 \left(\alpha +Q^2\right)}{2l^2 \sqrt{2\,\alpha \mathcal{W}}} \;\;;\;\;
    T = \frac{2 \alpha  \mathcal{W}\left(l^2+6 \alpha  (\varepsilon +1) \mathcal{W}\right)-l^2 \left(\alpha +Q^2\right)}{8 l^2 \sqrt{2} \pi  (\alpha\mathcal{W})^{3/2}} \ . 
\end{eqnarray}
Using the Eq.~\eqref{Perturbed EGB}, we can compute the perturbation parameter $\varepsilon$ as 
\begin{equation}
    \varepsilon =  -\frac{l^2 \left(\alpha-2\,M\, \sqrt{2\,\alpha\,\mathcal{W}}+2 \alpha  \mathcal{W}+Q^2\right)}{4 \alpha ^2 \mathcal{W}^2}-1 \ .
\end{equation}
Using this, we can compute the 
\begin{equation}\label{Tds for EGB}
    T\, \left(\frac{\partial S}{\partial \varepsilon}\right)_{M_{ext}} = -\frac{\sqrt{2} \alpha ^{3/2} \sqrt{\mathcal{W}} \left(\mathcal{W}+1\right)}{l^2} \ .
\end{equation}
Finally, to test whether the universality relation is satisfied or not, we have to differentiate the mass as in Eq.~\eqref{Perturbed EGB} w.r.t $\varepsilon$ and result in 
\begin{equation}\label{DM DE for EGB}
    \left(\frac{\partial M_{ext}}{\partial \varepsilon}\right) = \frac{\sqrt{2} \alpha ^{3/2} \mathcal{W}^{3/2}}{l^2} \ .
\end{equation}
To verify the Generalized uncertainty relation~\eqref{Generalized Universal Relation}, the last quantity needs to be computed using the horizon radius~\eqref{EGB Horizon} are
\begin{eqnarray}\label{RHS EGB gener.}
    ^{EGB}S_{BH}^h &=&  2\pi\;\alpha \, W\left(\frac{\ell_0^2\, e^{\frac{S}{2 \, \pi \, \alpha }}}{2 \alpha }\right) = 2\pi\;\alpha \,\mathcal{W}  \nonumber \\ 
    \frac{\partial (^{EGB}S_{BH}^h)}{\partial S}&=&\frac{W\left(\frac{p^2 e^{\frac{S}{2 \pi  \alpha }}}{2 \alpha }\right)}{W\left(\frac{p^2 e^{\frac{S}{2 \pi  \alpha }}}{2 \alpha }\right)+1} = \frac{\mathcal{W}}{\mathcal{W}+1}
\end{eqnarray}
Finally, we will verify the generalized universal relation Eq.~\eqref{Generalized Universal Relation}, by computing L.H.S. using Eq.~\eqref{Tds for EGB} and Eq.~\eqref{DM DE for EGB} and R.H.S. as computed in Eq.~\eqref{RHS EGB gener.}; it is easy to verify 
\begin{equation}
  -  \frac{(\partial M_{ext}/\partial \varepsilon)}{T\,(\partial S/\partial \varepsilon)_{M_{ext}}}  = \frac{\partial (^{EGB}S_{BH}^h)}{\partial S} \ .
\end{equation}
The generalized universal relation is satisfied. Now we can discuss a similar case with entropy correction, such as logarithmic and exponential, and verify the Generalized universal relation.

\section{Logarithmic correction}\label{Sec:Logarithmic correction}

The emergence of logarithmic corrections to the Bekenstein-Hawking entropy, which, in the semiclassical limit, scales as the horizon area. While this area law captures the leading behavior, it fails to address the quantum microstates underpinning the entropy, especially in non-extremal regimes where full microscopic control remains elusive. String theory, as a candidate for ultraviolet completion of gravity, has provided a compelling framework in which to probe the microscopic structure of black hole entropy. For certain classes of extremal and near-extremal black holes—often embedded in supersymmetric compactifications—exact microstate counting becomes tractable, yielding a statistical entropy that precisely matches the semiclassical result. These extremal configurations, characterized by the degeneracy of inner and outer horizons and vanishing Hawking temperature, exhibit enhanced near-horizon symmetries, often governed by an emergent AdS\(_2\) throat geometry. This deepened symmetry structure plays a pivotal role in both holographic duality and entropy counting, linking black hole thermodynamics to two-dimensional conformal field theories. In contrast, non-extremal black holes—those with non-degenerate horizons and non-zero surface gravity—pose significant challenges. Their thermal behavior implies a dynamical flow towards extremality as they radiate away energy. However, in such cases, the lack of a supersymmetric or protected structure obstructs a first-principles statistical derivation of entropy, highlighting the limitations of our current non-perturbative understanding of quantum gravity. Beyond the leading order, corrections to the Bekenstein–Hawking entropy carry the imprint of quantum gravitational effects. These subleading terms, particularly the logarithmic corrections, have been computed using various approaches: Euclidean quantum gravity, loop quantum gravity, and string-theoretic modular invariance techniques. Generically, the corrected entropy can be expressed as:
\[
S_{\text{BH}} = \frac{A_H}{4 G_N} + \alpha \ln \left( \frac{A_H}{G_N} \right) + \cdots \ .
\]
where \( \alpha \) encodes the one-loop quantum correction arising from massless field fluctuations around the black hole background. The logarithmic term is universal in many setups, sensitive only to low-energy data, while the power-law corrections reflect deeper quantum gravitational structure. Despite these advances, a unified microscopic derivation encompassing both extremal and non-extremal black holes remains incomplete. The general form for the Logarithmic correction is 
\begin{equation}\label{Log Horizon}
    S_{\ell} = \alpha_{\ell}\, r_h^2 + \beta_{\ell}\, \ln\left({\gamma_{\ell} \, r_h^2}\right) \ ,
\end{equation}
where $r_h$ is the horizon radius and $\alpha_{e}$, $\beta_{e}$ and $\gamma_{e}$ are constants. In the terms of $S_{e}$, the horizon radius is 
\begin{equation}\label{Horizon Radius Log}
    r_h =  \sqrt{\frac{\beta_{\ell} }{\alpha_{\ell}} \,\mathcal{W_L}} \ .
\end{equation}
Here we have defined $W\left(\frac{\alpha_{\ell}  e^{S_{\ell}/\beta_{\ell} }}{\beta_{\ell}  \gamma_{\ell} }\right) =\mathcal{W_L}.$
By applying this framework and perturbing the action in a manner analogous to Eq.~\eqref{Perturbation Action}, the mass receives quantum corrections, leading to the modified expression of mass 
\begin{eqnarray}\label{Dm by d epsilon Log}
    M_L(\varepsilon) &=& \frac{\alpha_{\ell} ^2 l^2 Q^2+\beta_{\ell}  \mathcal{W_L} \left(\alpha_{\ell}  l^2+\beta_{\ell}  (\varepsilon +1) \mathcal{W_L}\right)}{2 \alpha_{\ell} ^{3/2}  l^2 \sqrt{\beta_{\ell} \,\mathcal{W_L}}} \ .
    \end{eqnarray}
Upon solving the above relation, the perturbation parameter \( \varepsilon \) can be readily determined as
\begin{equation}\label{Log Exponential}
    \varepsilon =   -\frac{\alpha_{\ell}  l^2 \left(-2 \sqrt{\alpha_{\ell} \, \beta_{\ell} } M \sqrt{\mathcal{W_L}}+\alpha_{\ell}  Q^2+\beta_{\ell}  \mathcal{W_L}\right)}{\beta_{\ell} ^2 \mathcal{W_L}^2}-1  \ .
\end{equation}
The corrected temperature is 
    \begin{eqnarray}\label{TLog Epsilon}
    T_L(\varepsilon) &=& \frac{\beta_{\ell}  \mathcal{W_L} \left(\alpha_{\ell}  l^2+3 \beta_{\ell}  (\varepsilon +1) \mathcal{W_L}\right)-\alpha_{\ell} ^2 l^2 Q^2}{4 l^2 \pi  \sqrt{\alpha_{\ell} } \beta_{\ell} ^{3/2} \mathcal{W_L}^{3/2}}
\end{eqnarray}
The final quantity, namely the denominator of Eq.~\eqref{Generalized Universal Relation}, is given by 
\begin{eqnarray}\label{Tds by d epsilon Log}
    T_L(\varepsilon)\, \left(\frac{\partial \, S_{\ell}}{\partial \, \varepsilon}\right)\Bigg|_{M_{ext}} = -\frac{\beta_{\ell}^{3/2} \sqrt{\mathcal{W_L}} \left(\mathcal{W_L}+1\right)}{2 \, \pi  \sqrt{\alpha_{\ell} } \ell^2} \ .
\end{eqnarray}
We now turn to the verification of the generalized universal relation expressed in Eq.~\eqref{Generalized Universal Relation}, within the context of general logarithmic correction in the entropy.  The right-hand side is evaluated using Eq.~\eqref{Horizon Radius Log} as
\begin{eqnarray}\label{RHS of Gen Ent. Log}
    \frac{\partial \left(^{Log.}S_{BH}^h\right)}{\partial \, S_{\ell}} = \frac{\pi \mathcal{W_L}}{\alpha_{\ell}  \left(\mathcal{W_L}+1\right)} \ .
\end{eqnarray}
Finally, we compute the left-hand side using the thermodynamic identities for the extremal mass and entropy derivatives, as given in Eq.~\eqref{Dm by d epsilon Log}, Eq.~\eqref{Tds by d epsilon Log}, and Eq.~\eqref{RHS of Gen Ent. Log}, respectively, one can verify
\[
-\frac{\left( \partial M_{\text{ext}}/\partial \varepsilon \right)}{T\,\left(\partial S_{\ell}/\partial \varepsilon \right)_{M_{\text{ext}}}}  = \frac{\partial (^{Log.}S_{BH}^h)}{\partial S} \,,
\]
thereby confirming the validity of the generalized universal relation in the extremal limit of logarithmically corrected entropy.
\section{Exponential correction}\label{Sec:Exponential COrrection}

The origin of exponential correction comes from assuming a black hole composed of \( N \) fundamental particles. From statistical mechanics, the entropy of such a system can be derived by calculating the total number of microstates it can occupy. If the system is described by occupation numbers \( n_i \), shared among \( s_i \) identical particles in different states, the number of microstates depends on the combinations of these distributions. Each microstate is associated with an energy \( \varepsilon_i \), and the total number of particles and the total energy are given by summing over all states. In the limit of large \( N \), Stirling’s approximation allows us to simplify the expressions for the number of microstates. By maximizing the entropy under the constraints of fixed total particle number and energy, we derive the most probable distribution of particles among the microstates. This leads to an exponential form involving a parameter \( \lambda \), which acts as a Lagrange multiplier. Upon normalization, \( \lambda \) is found to be approximately \( \ln 2 \), with small corrections that decay exponentially with increasing \( N \). Using this, the entropy of the black hole is shown to be proportional to \( N \), and after eliminating \( N \) in favor of the classical entropy \( S_0 \), one can find that the quantum-corrected entropy takes exponential correction to the entropy as \( S = S_0 + e^{-S_0} \). Here, \( S_0 = A / (4 l_p^2) \) is the standard Bekenstein–Hawking entropy, \( A \) is the horizon area, and \( l_p \) is the Planck length. This result indicates an exponential correction to black hole entropy, which becomes significant at small scales and vanishes for large horizon areas. Following~\cite{Chatterjee:2020iuf}, the general form for the Exponential correction is 
\begin{equation}\label{Horizon Exponential}
    S_{e} = \alpha_{e}\, r_h^2 + \beta_{e}\, e^{\gamma_{e} \, r_h^2} \ ,
\end{equation}
where $r_h$ is the horizon radius and $\alpha_{e}$, $\beta_{e}$ and $\gamma_{e}$ are constants. In the terms of $S_{e}$, the horizon radius is 
\begin{equation}\label{Horizon Radius Exponential}
    r_h^2 =  \frac{S_{e}}{\alpha_{e}}-\frac{\mathcal{W_E}}{\gamma_{e}}  \ .
\end{equation}
Here we have defined $\mathcal{W_E}= W\left(\frac{\beta_{e}\,  \gamma_{e}\,  e^{\frac{\gamma_{e}\,  S_{e}}{\alpha_{e} }}}{\alpha_{e} }\right)$. With the help of this horizon radius, the Bekenstein-Hawking entropy and its derivative with respect to entropy is 
\begin{eqnarray}\label{RHS Exp gener.}
    ^{Exp.}S_{BH}^h &=& \frac{\pi  S}{\alpha_e }-\frac{\pi  \mathcal{W_E}}{\gamma_e } \;\;\;\;\;;\;\;\;\;\;\frac{\partial \left( ^{Exp.}S_{BH}^h \right)}{\partial S} = \frac{\pi }{\alpha_e +\alpha_e  \mathcal{W_E}}
\end{eqnarray}

By using this and perturbing the action similar to \eqref{Perturbation Action}, the mass gets corrected, and we have 
\begin{eqnarray}\label{Mass in Exponential}
    M(\varepsilon) = \frac{\gamma_{e}^2 \left(\alpha_{e} ^2 l^2 Q^2+\alpha_{e}  l^2 S_{e}+S_{e}^2 (\varepsilon +1)\right)+\alpha_{e}  \mathcal{W_E} \left(\alpha_{e}  (\varepsilon +1) \mathcal{W_E}-\gamma_{e}  \left(\alpha_{e}  l^2+2 S_{e} (\varepsilon +1)\right)\right)}{2l^2 \alpha_{e} ^{3/2} \gamma_{e} ^{3/2}  \sqrt{\gamma_{e}  S_{e}-\alpha_{e}  \mathcal{W_E}}} \ ,
    \end{eqnarray}
and by solving this, we can easily compute the perturbation parameter $\varepsilon$ as 
\begin{equation}\label{Epsilon Exponential}
    \varepsilon =   \frac{\alpha_{e}  \gamma_{e}  l^2 \left(2 \sqrt{\alpha_{e} \,\gamma_{e} } M \sqrt{\gamma  S_{e}-\alpha_{e}  \mathcal{W_E}}-\left(\gamma_{e}  \left(\alpha_{e}  Q^2+S_{e}\right)\right)+\alpha_{e}  \mathcal{W_E}\right)}{\left(\gamma_{e}  S_{e}-\alpha_{e}  \mathcal{W_E}\right)^2}-1  \ . \nonumber
\end{equation}
The corrected temperature of the black hole is 
\begin{eqnarray}\label{TEXP Epsilon}
    T(\varepsilon) = \frac{\gamma_{e} ^2 \left(\alpha_{e} ^2 l^2 \left(-Q^2\right)+\alpha_{e}  l^2 S_{e}+3 S_{e}^2 (\varepsilon +1)\right)+\alpha_{e}  \mathcal{W_E} \left(3 \alpha_{e}  (\varepsilon +1) \mathcal{W_E}-\gamma_{e}  \left(\alpha_{e}  l^2+6 S_{e} (\varepsilon +1)\right)\right)}{4 \pi  \sqrt{\alpha_{e} \, \gamma_{e} } l^2 \left(\gamma_{e}  S_{e}-\alpha_{e}  \mathcal{W_E}\right)^{3/2}} \ . \nonumber
\end{eqnarray}
The last quantity, i.e., the denominator of Eq.~\eqref{Generalized Universal Relation}, is  
\begin{eqnarray}\label{Tds by d epsilon Exponential}
    T(\varepsilon)\, \left(\frac{\partial \, S_{e}}{\partial \, \varepsilon}\right)\Bigg|_{M_{ext}} = -\frac{\left(\mathcal{W_E}+1\right) \left(\gamma_{e}  S_{e}-\alpha_{e}  \mathcal{W_E}\right)^{3/2}}{2 \pi  \sqrt{\alpha_{e} } \gamma_{e} ^{3/2} l^2} \ .
\end{eqnarray}
Finally, we proceed to verify the validity of the generalized universal relation given in Eq.~\eqref{Generalized Universal Relation}. This is achieved by evaluating its left-hand side (L.H.S.) using the thermodynamic expressions provided in Eq.~\eqref{Mass in Exponential} and Eq.~\eqref{Tds by d epsilon Exponential}, and using the right-hand side (R.H.S.) using Eq.~\eqref{RHS Exp gener.}. It is easy to verify
\[
-\frac{\left( \partial M_{\text{ext}}/\partial \varepsilon \right)}{T\,\left(\partial S/\partial \varepsilon \right)_{M_{\text{ext}}}} =\frac{\partial (^{Exp.}S_{BH}^h)}{\partial S} \ ,
\]
demonstrating that the generalized universal relation is indeed satisfied within this framework.

\section{Summary}\label{Sec:Conclusion and Discussion}

In summary, we have derived a generalized universal relation based on Goon and Penco's universal relation. In particular, we assumed the entropy is not just the Bekenstein-Hawking entropy; there should be some correction to that. We have studied perturbative as well as non-perturbative corrections to the Bekenstein-Hawking entropy. The perturbative correction corrects the entropy by a logarithmic correction; at the same time, the non-perturbative corrections add an exponential correction to the Bekenstein-Hawking entropy. We have shown that the universal relation proposed by Goon and Penco is not satisfied whenever the entropy gets an extra correction, either due to the perturbative correction or the non-perturbative correction. It is worth looking at those situations where the entropy is just the Bekenstein-Hawking entropy, but has some other form.

\quad We have first derived the generalized form of the Universal relation and called it the Generalized Universal relation. The word generalized means we have started with the entropy as a function of the horizon radius. This function covers the Bekenstein-Hawking entropy as well as all the possible corrections to the entropy. Since it is studied in the literature that the EGB gravity gets a correction in the metric function due to the EGB parameter, and then a logarithmic correction in the entropy. Using that fact, we have checked our derived relation in such a case and showed that the derived Generalized Universal Relation is satisfied. After verifying the derived relation, we delve into a quantum correction to the Bekenstein-Hawking entropy, departing from the classical Bekenstein–Hawking entropy. 

\quad We consider leading-order quantum modifications in the form of (i) logarithmic corrections, typically arising from one-loop quantum fluctuations, and (ii) exponential corrections, which may be motivated by specific quantum gravity proposals. These corrections are incorporated perturbatively by modifying the entropy functional while maintaining the background governed by the Einstein–Hilbert–Maxwell action. Since exponential correction arises naturally in certain models involving non-perturbative quantum gravity effects, logarithmic correction is well-motivated by a wide class of quantum gravity approaches, including loop quantum gravity and string-theoretic microstate counting. In both cases, the corrected entropy functions are used to compute the associated thermodynamic quantities, including the corrected Hawking temperature and mass in the extremal limit. Using that, we derived the horizon radius and the perturbed thermodynamic quantities. We explicitly evaluate both sides of the generalized universal relation by utilizing these corrected quantities. Despite the presence of subleading quantum corrections, we find that the relation remains satisfied to the appropriate perturbative order. This provides strong evidence that the generalized universal relation possesses a degree of universality that extends beyond the classical regime and remains valid under a broad class of quantum corrections to black hole entropy.

\quad It would be interesting to investigate and verify the Generalized Universal relations in systems where entropy is not just the Bekenstein-Hawking entropy. One should consider the Horava-Lifshitz black hole~\cite{Pourhassan:2022nay}, the magnetic charge AdS black holes~\cite{Gogoi:2024ypn}, STU supergravity models \cite{Karan:2024gwf}, modified gravity theories \cite{Pourhassan:2024yfg, Soroushfar:2023acx}. We will hope to return to this issue elsewhere.

\acknowledgments

A.A. is financially supported by the IIT Kanpur Institute’s postdoctoral fellowship. I want to thank Aditya Singh (IIT Dhanbad) for helpful comments on the draft.


\appendix

 \section{Lambert W function}\label{Appn:Lambert W function}

The \textbf{Lambert W function}, also known as the omega function or product logarithm, is defined as the set of functions \( W(z) \) that satisfy the equation
\[
z = W(z) \cdot e^{W(z)}
\]
for any complex number \( z \). In other words, it is the inverse of the function \( f(w) = w e^w \). The function is \textit{multivalued}, with infinitely many branches indexed by integers \( k \). The principal branch (\( W_0 \)) is real for \( z \geq -\frac{1}{e} \). The secondary branch (\( W_{-1} \)) is real for \( -\frac{1}{e} < z < 0 \). For real arguments, only \( W_0 \) and \( W_{-1} \) are typically relevant. The Lambert W function cannot be expressed in terms of elementary functions. All branches except \( W_0 \) have a logarithmic singularity at \( z = 0 \).

\quad Example:
\begin{itemize}
    \item Consider the equation \( x e^x = 5 \). The solution can be expressed as
\[
x = W(5)
\]
This provides the value of \( x \) that satisfies the equation.
\item  To solution of $x = a + b e^{c x}$ is
\[
x = a - b \cdot W\left(-\frac{1}{bc} e^{-ca} \right)
\]

\end{itemize}

\section{$S=\pi r_h^\mathrm{a}$ universal Relation}\label{Appn:rh^a Entropy}

In the case of this entropy relation, the corrected mass and temperature are 
\begin{eqnarray}
    M(\varepsilon) &=& \frac{1}{2} \left(\pi^3 S\right)^{-1/\mathrm{a}} \left(\frac{(\varepsilon +1) S^{4/\mathrm{a}}}{l^2}+\pi ^{4/\mathrm{a}} Q^2+\pi ^{2/\mathrm{a}} S^{2/\mathrm{a}}\right)\\
    T(\varepsilon) &=& \frac{1}{4} \pi ^{-\frac{\mathrm{a}+1}{\mathrm{a}}} S^{-3/\mathrm{a}} \left(\frac{3 (\varepsilon +1) S^{4/\mathrm{a}}}{l^2}-\pi ^{4/\mathrm{a}} Q^2+\pi ^{2/\mathrm{a}} S^{2/\mathrm{a}}\right) \ .
\end{eqnarray}
Using this, we can compute the perturbation parameter $\varepsilon$ as 
\begin{eqnarray}\label{rha Epsilon}
    \varepsilon = \pi ^{2/\mathrm{a}} \left(-l^2\right) S^{-4/\mathrm{a}} \left(-2 \pi ^{1/\mathrm{a}} M S^{1/\mathrm{a}}+\pi ^{2/\mathrm{a}} Q^2+S^{2/\mathrm{a}}\right)-1 \ .
\end{eqnarray}
Now, we can compute the numerator and denominator of the universal relation \eqref{Universal Relation} as
\begin{eqnarray}
    \frac{\partial M_{ext}}{\partial \varepsilon} &=& \frac{\pi ^{-3/\mathrm{a}} S^{3/\mathrm{a}}}{2 l^2}\\
   T(\varepsilon) \left(\frac{\partial S}{\partial \varepsilon}\right)\Bigg|_{M_{ext}} &=& \frac{\mathrm{a} \pi ^{-\frac{\mathrm{a}+3}{\mathrm{a}}} S^{\frac{1}{\mathrm{a}}+1} \left(\pi ^{2/\mathrm{a}} l^2 \left(S^{2/\mathrm{a}}-\pi ^{2/\mathrm{a}} Q^2\right)+3 (\varepsilon +1) S^{4/\mathrm{a}}\right)}{8 l^4 \left(-3 \pi ^{1/\mathrm{a}} M S^{1/\mathrm{a}}+2 \pi ^{2/\mathrm{a}} Q^2+S^{2/\mathrm{a}}\right)} \ .
\end{eqnarray}
Finally, using Eq.~\eqref{rha Epsilon}, we have 
\begin{equation}
     -\,\frac{\left(\partial M_{ext}/\partial\varepsilon \right)}{ T \left(\partial S/\partial\varepsilon \right)|_{M_{ext}}} = \frac{2}{\mathrm{a}} \left(\frac{S}{\pi}\right)^{\frac{2-\mathrm{a}}{\mathrm{a}}} \ .
\end{equation}
It is easy to see that the universal relation is verified for $\mathrm{a}=2$; otherwise, it is not. 


\bibliographystyle{JHEP}
\bibliography{References}

\end{document}